\documentclass[9pt,conference]{IEEEtran}
\IEEEoverridecommandlockouts
\usepackage{amssymb,amsthm,amsmath,array}
\usepackage{graphicx}
\usepackage{xspace}
\usepackage[sort&compress, numbers]{natbib}
\usepackage{stmaryrd}
\usepackage{xcolor}
\usepackage{mathtools}
\usepackage{float}
\usepackage{textcomp}
\usepackage{mathdefs}
\usepackage{enumitem}
\usepackage{titlesec}
\usepackage{siunitx}

\titlespacing*{\section}{0pt}{1pt}{1pt}
\titlespacing*{\subsection}{0pt}{1pt}{1pt}

\usepackage[labelformat=simple]{subcaption}

\begin{document}
\title{Indoor Airflow Imaging Using \\Physics-Informed Background-Oriented Schlieren Tomography \thanks{$^1$Arjun Teh conducted this work during an internship at MERL}}
\author{\IEEEauthorblockN{
        Arjun Teh\IEEEauthorrefmark{1}$^1$, 
        Wael H. Ali\IEEEauthorrefmark{2},
        Joshua Rapp\IEEEauthorrefmark{2}, and 
        Hassan Mansour\IEEEauthorrefmark{2}
    }
    \IEEEauthorblockA{
        \IEEEauthorrefmark{1} Carnegie Mellon University, Pittsburgh, PA, USA\\
        \IEEEauthorrefmark{2} Mitsubishi Electric Research Laboratories, Cambridge, MA, USA}
}
\maketitle
\begin{abstract}
We develop a framework for non-invasive volumetric indoor airflow estimation from a single viewpoint using background-oriented schlieren (BOS) measurements and physics-informed reconstruction. 
Our framework utilizes a light projector that projects a pattern onto a target back-wall and a camera that observes small distortions in the light pattern.
While the single-view BOS tomography problem is severely ill-posed, our proposed framework addresses this using: (1) improved ray tracing, (2) a physics-based light rendering approach and loss formulation, and (3) a physics-based regularization using a physics-informed neural network (PINN) to ensure that the reconstructed airflow is consistent with the governing equations for buoyancy-driven flows.
\end{abstract}

\section{Introduction}
Understanding the airflow in indoor spaces is crucial for improving the comfort and efficiency of heating, ventilation, and air conditioning (HVAC) systems~\cite{patankar_airflow_2010, abedi_smart_2018}.
However, three-dimensional (3D) airflow sensing is challenging since hardware sensors only measure localized spatial regions around the sensors~\cite{okamoto_digital_1994}.

One promising imaging technique is background-oriented schlieren (BOS) tomography, which uses images of a patterned background to observe distortions due to changes in the refractive index of a transparent medium~\cite{raffel_applicability_2000, dalziel_whole-field_2000, meier_computerized_2002,Settles2017ARO,settles_BOSvelocimetry_2018}.
While BOS has been effective for quantitative measurement of gas flows with high spatial resolution, 
the tomographic inverse problem is only well-posed with a sufficient number of view angles~\cite{atcheson_time-resolved_2008, nicolas_direct_2015, grauer_instantaneous_2018, cai_flow_2021}. 

In this work, we combine several recent advances in refractive field tomography to achieve 3D airflow reconstructions for room-scale scenes from a single camera view. 
First, we use a physics-informed neural network (PINN) framework~\cite{raissi_physics-informed_2019} as a regularizer to ensure that the reconstructed field adheres to the partial differential equations (PDEs) governing airflow. This approach is similar to those suggested in~\cite{zhao_single_2024, molnar_estimating_2023, molnar_physics-informed_2023, molnar_forward_2024, cai_flow_2021} but applied for the first time in a single view room-scale setting. 
Additionally, we employ improved ray tracing and use a physics-based rendering approach according to the refractive radiative transfer equation (RRTE)~\cite{ament_refractive_2014, pediredla_path_2020}.
We also explore the use of a light source projecting a pattern onto the wall~\cite{weisberger_projection_2022}, which may be more practical than a fixed background.
This extended abstract summarizes our approach, with a comprehensive version of this work~\cite{teh2025IndoorPub} set to appear in the proceedings of ICASSP 2025.

\section{BOS Imaging Formulation}
\subsection{Airflow Imaging Setup}
We consider a BOS imaging scenario comprising an air-filled room, a camera, and either a patterned background wall or a light source that can project a pattern on the back-wall, as shown in Fig.~\ref{fig:gt}. When no air is flowing, the camera captures a reference image $I_{\rm ref}$ of the back-wall pattern. When the inlet on the side wall blows the airflow into the room, the captured image $I_{\rm flow}$ appears distorted due to the change in density of the air which induces a gradient in its refractive index $\eta$, 
as described by the Gladstone--Dale equation 
    $\eta = 1 + G\rho$,
where $G$ is the Gladstone--Dale coefficient~\cite{gladstone_xiv_1863}.
Assuming the pressure variation of air in the room is small, then by the ideal gas law, the refractive index is related to temperature $T$ as $\eta(T) = 1 + \rho_0 G\frac{T_0}{T}$,
where $\rho_0$ is the ambient density and $T_0$ is the ambient temperature~\cite{cai_flow_2021}. It can be seen that changes in the air temperature cause changes in the refractive index causing light rays passing through the air to bend.

\subsection{Ray Tracing Formalism}
The propagation of light inside a medium with continuously varying refractive index can be described using the ray tracing ordinary differential equations (ODEs)~\cite{sharma1982Tracing}
\begin{equation}
\begin{array}{ll}
    \dfrac{\mathrm{d}\pos\left(t\right)}{\mathrm{d}t} = \vel\left(\pos\left(t\right)\right)\,, &\
    \dfrac{\mathrm{d}\vel\left(t\right)}{\mathrm{d}t} = \eta\left(\pos\left(t\right)\right) \nabla \eta\left(\pos\left(t\right)\right). \label{eqn:raytracevel}
\end{array}
\end{equation}
Due to the implicit dependence of the refractive index $\eta$ on the ray position $\pos\left(t\right)$, the ODE system is fully coupled and its solution is referred to as \textit{nonlinear ray tracing}. The fully coupled nature of this ODE system introduces significant computational complexity. To address this, we employ a \textit{quasi-linear ray tracing} approach to decouple the velocity ODE from the position ODE, simplifying the computation with little deviation from the original, true path. Specifically, since the change in refractive index is small, we can approximate the trajectory of the light ray as a straight line starting from $\pos_0$ along $\vel_0$. The refractive index field is queried along this linear path to compute an approximate velocity $\Tilde\vel(t)$ that is then used in the position ODE to obtain the updated ray path $\Tilde\pos(t)$. 

\subsection{Image Formation Model}
We use a physics-based rendering approach -- specifically, the path integral expression of the RRTE~\cite{ament_refractive_2014,pediredla_path_2020} -- to compute the intensity $I_j$ for a given pixel $j$ on the sensor plane as
\begin{equation}
    I_j = \int_{A}\int_{\Omega} \camerafilter_j(\sensor{\pos}) \luminance_{\text{wall}}(\wall{\pos}, \wall{\vel})\frac{\left<\wall{\normal}, \wall{\vel} \right>}{\|r_{\rm s \leftrightarrow w}\|} \mathrm{d}\sensor{\vel}\mathrm{d}\sensor{\pos},
    \label{eqn:imageformation}
\end{equation}
where $\camerafilter_j$ is a triangular camera filter function, $(\wall{\pos}, \wall{\vel})$ are the wall position and velocity of a traced ray starting from $(\sensor{\pos}, \sensor{\vel})$ on the sensor plane, \(\luminance_{\text{wall}}\) is the luminance of the back-wall, $\left<\wall{\normal}, \wall{\vel} \right>$ is the cosine of the angle between the back-wall normal $\wall{\normal}$ and $\wall{\vel}$, and $\|r_{\rm s \leftrightarrow w}\|$ is the length of the ray path.
The total intensity for the pixel integrates over all starting ray velocities, $\sensor{\vel} \in \Omega$, and all starting ray positions within the area of the sensor pixel, $\sensor{\pos} \in A$. In practice, we use Monte Carlo sampling to evaluate this integral. 

To determine the luminance of the back-wall, we consider two cases. In the case where the back-wall is a textured light source, \(\luminance_{\text{wall}}\) is known and can be queried directly. If we model the wall as being illuminated by a pinhole projector, then a point on the back-wall will be illuminated by a single point from the projector. Further details regarding the back-wall-projector connection and the overall image formation model for this case can be found in~\cite{teh2025IndoorPub}.
\def\imHeight{2.9cm}
\begin{figure*}[h]
    \centering
    \begin{subfigure}{0.2\linewidth}
        \centering
        \caption{\vspace{-0.3ex} Ground Truth}
        \includegraphics[trim={3mm 1mm 1mm 17mm},clip, scale=0.19]{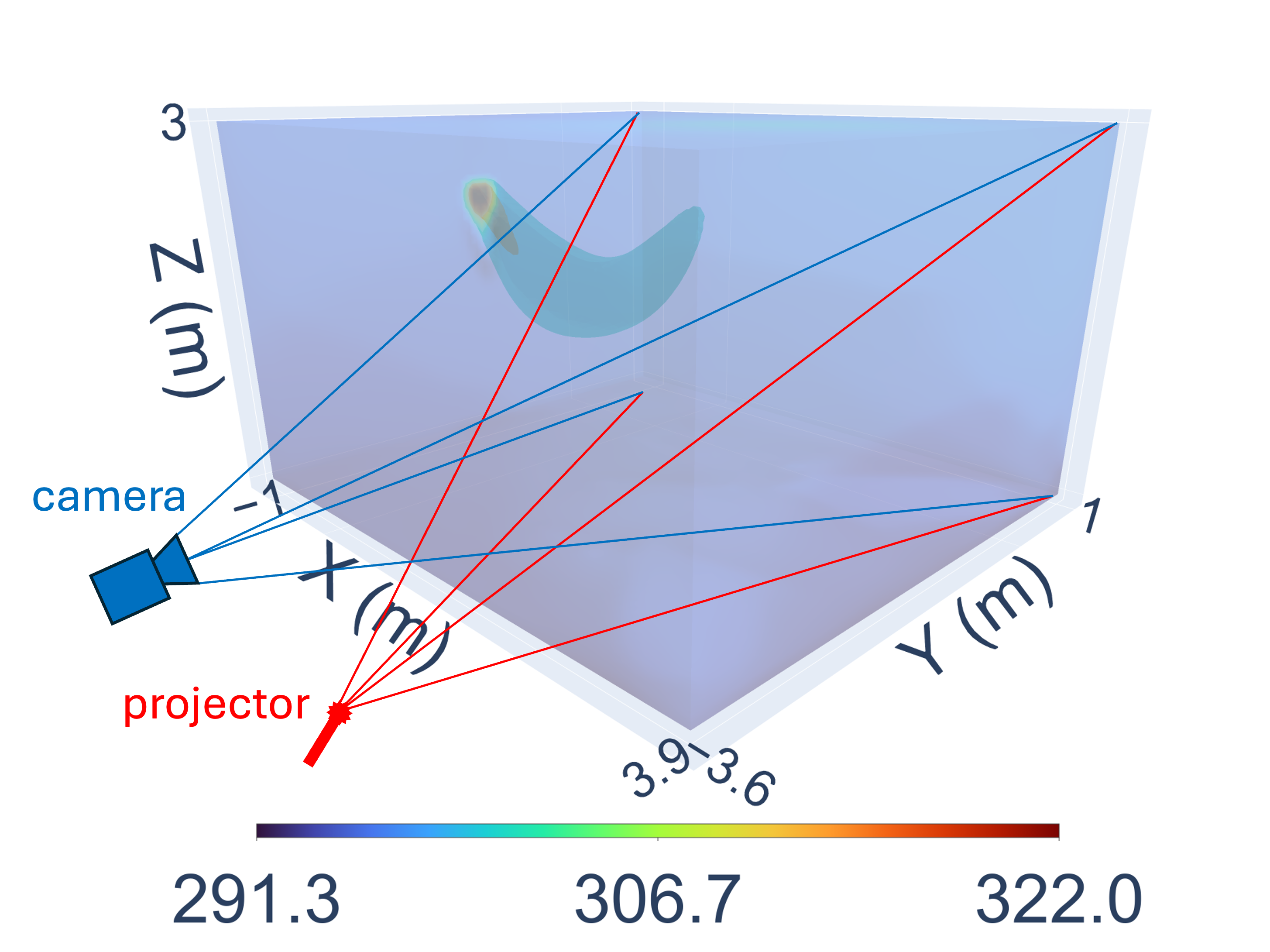}
        \label{fig:gt}
    \end{subfigure}
    \hfill
    \begin{subfigure}{0.02\linewidth}
        \caption*{\rotatebox[origin=l]{90}{\hspace{5em}Prediction}}
    \end{subfigure}
    \begin{subfigure}{0.2\linewidth}
        \centering
        \caption{\vspace{-0.3ex} BOS + Boundary}
        \includegraphics[trim={50mm 5mm 50mm 50mm},clip,height=\imHeight]{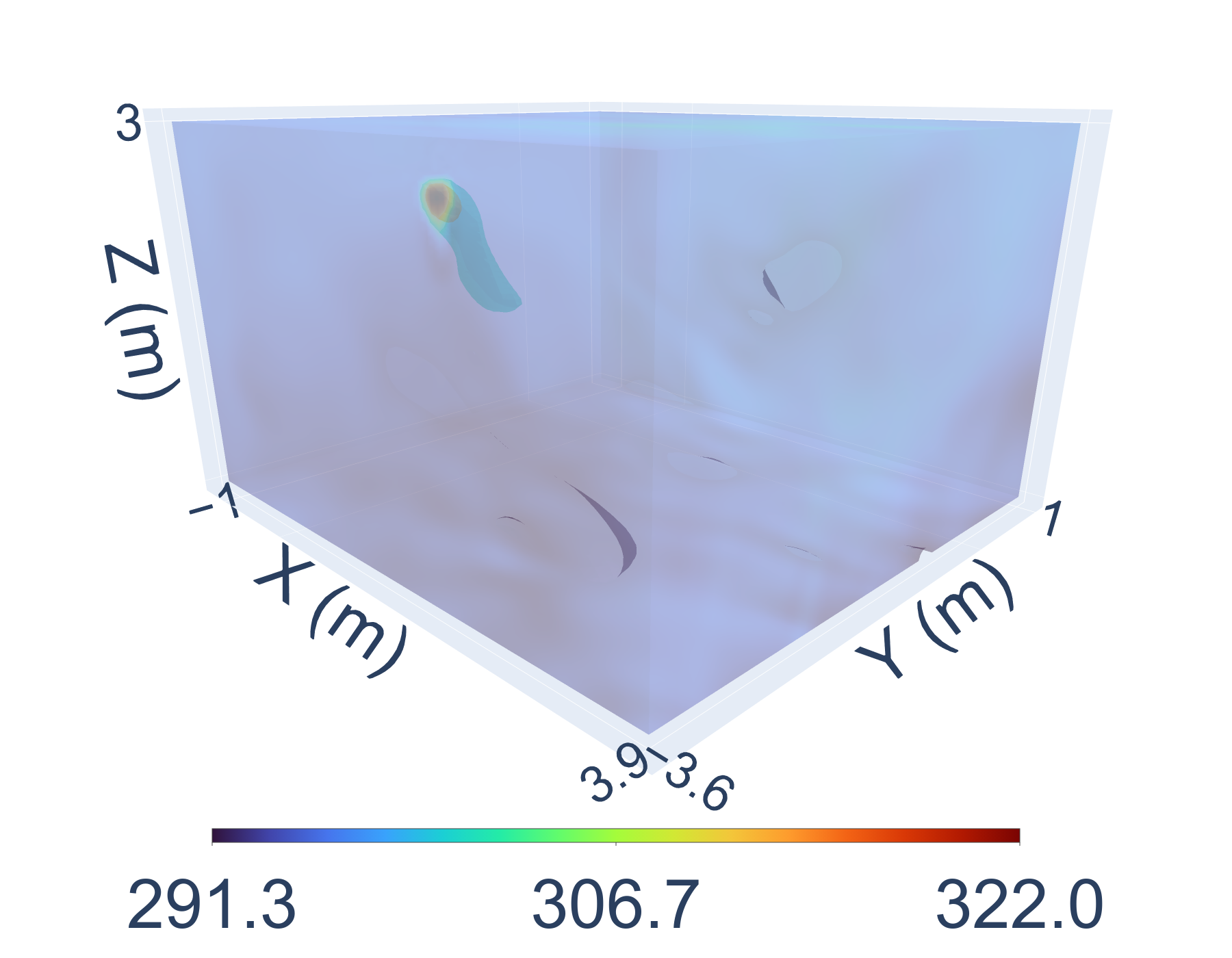}
        \label{fig:bos}
    \end{subfigure}
    \hfill
    \begin{subfigure}{0.2\linewidth}
        \centering
        \caption{\vspace{-0.3ex}PDE + Boundary}
        \includegraphics[trim={50mm 5mm 50mm 50mm},clip,height=\imHeight]{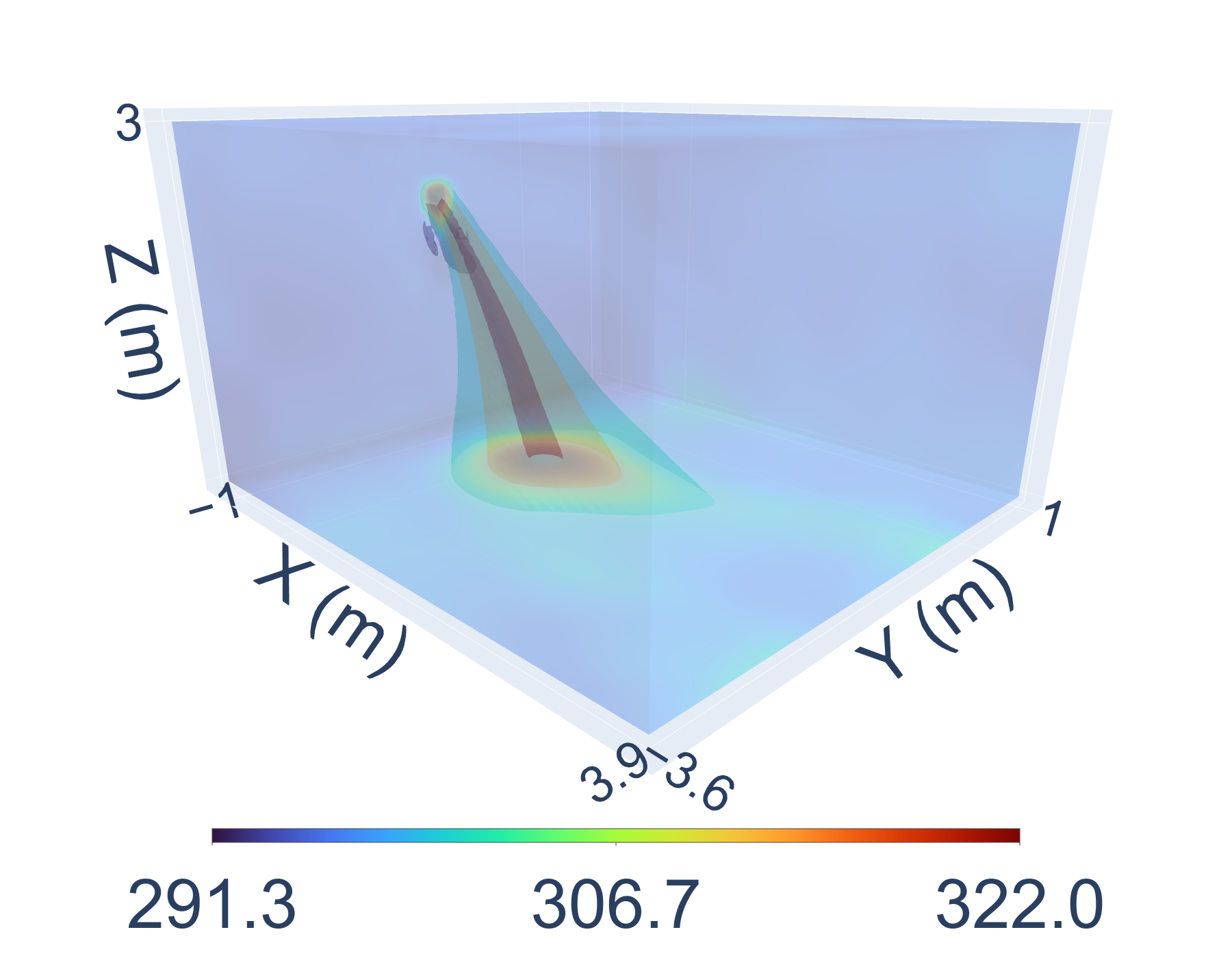}
        \label{fig:pde-boundary}
    \end{subfigure}
    \hfill
    \begin{subfigure}{0.2\linewidth}
        \centering
        \caption{\vspace{-0.3ex}BOS + PDE + Boundary}
        \includegraphics[trim={50mm 5mm 50mm 50mm},clip,height=\imHeight]{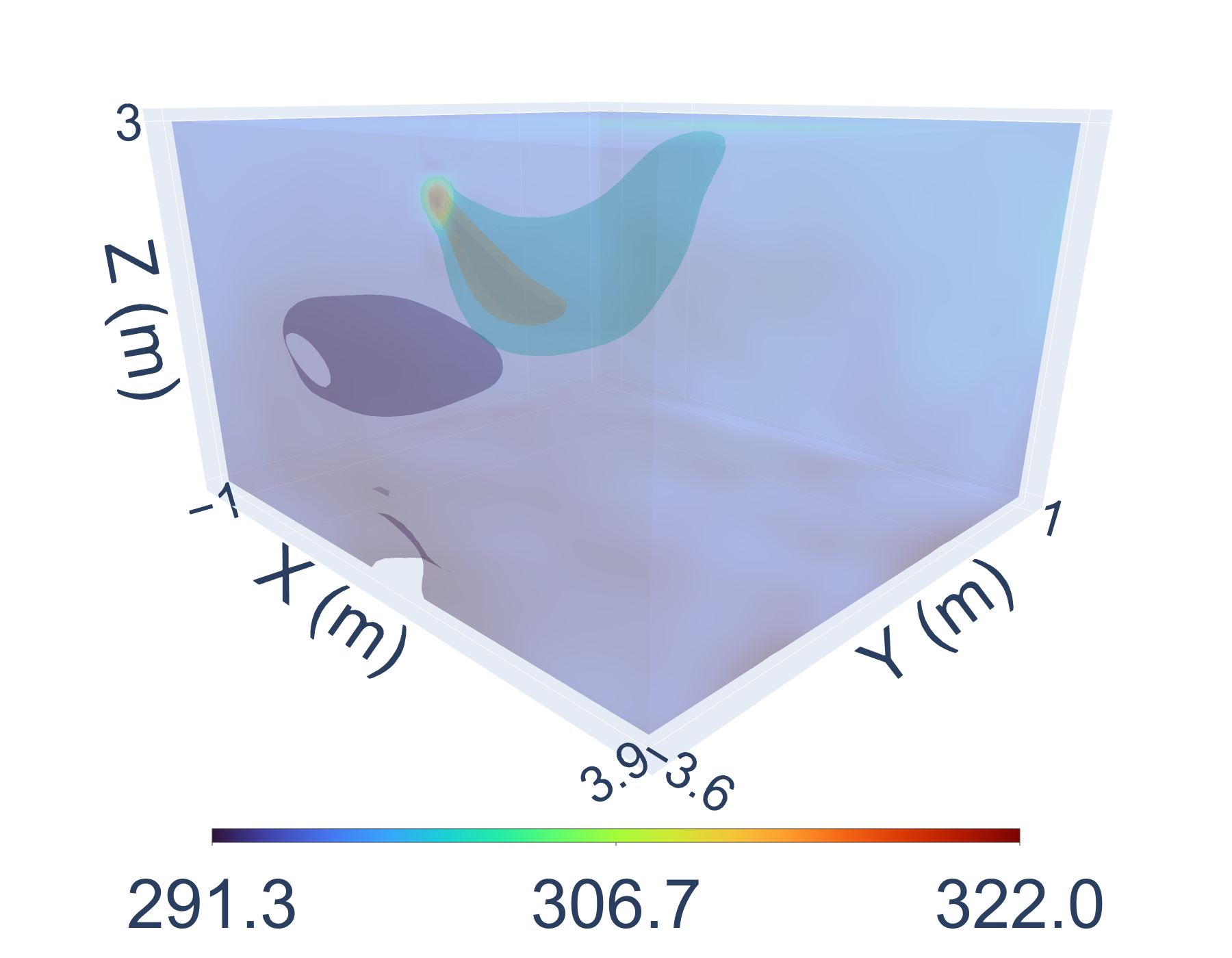}
        \label{fig:bos-pde-boundary}
    \end{subfigure}
    \\ \vspace{-1.5em}
    \begin{subfigure}{0.2\linewidth}
        \centering
        \caption{\vspace{-1.5ex} BOS Measurement}
        \includegraphics[trim={-18mm 0 0mm -8mm},clip,height=\imHeight]{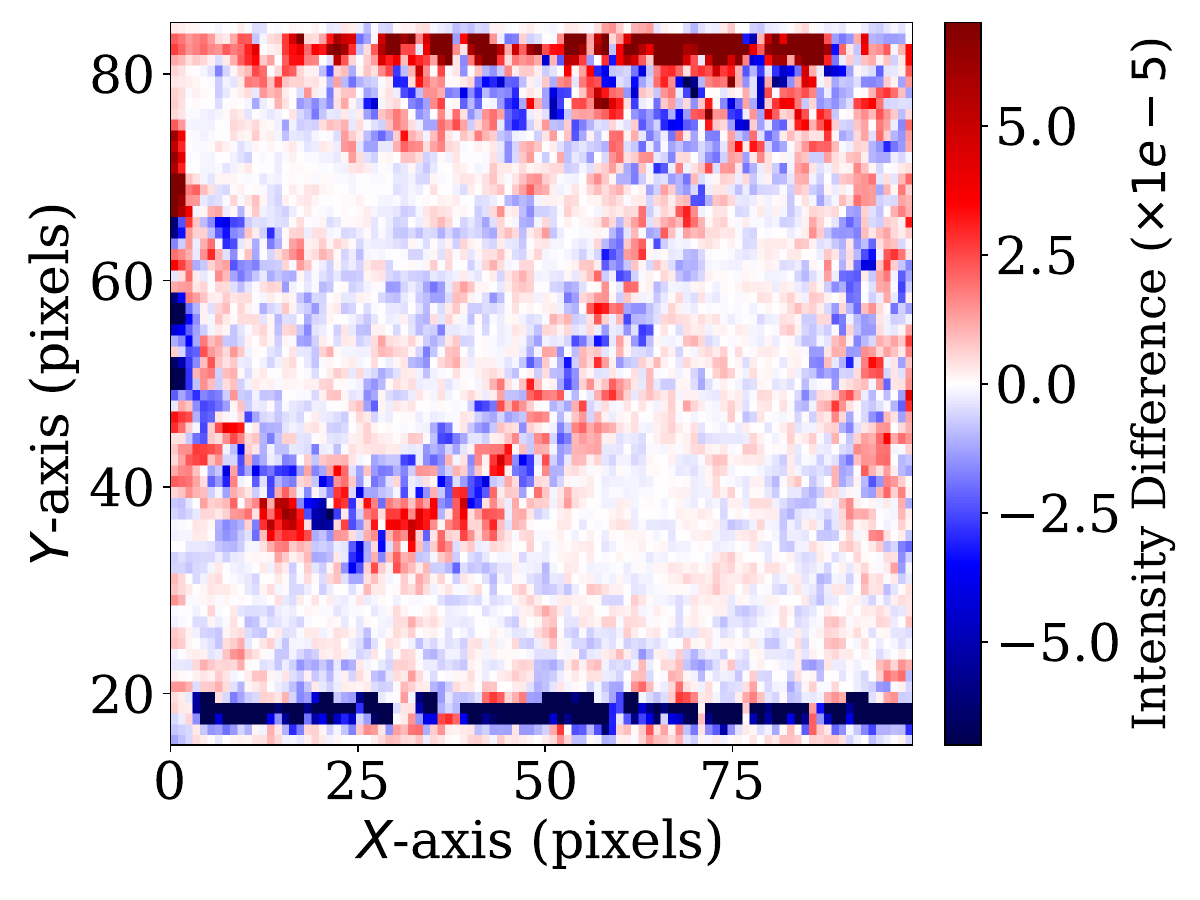}
        \label{fig:bos_meas}
    \end{subfigure}
    \hfill
    \begin{subfigure}{0.02\linewidth}
        \caption*{\rotatebox[origin=l]{90}{\hspace{6em} Error}}
    \end{subfigure}
    \begin{subfigure}{0.2\linewidth}
        \centering
        \caption*{~}
        \includegraphics[trim={50mm 5mm 50mm 50mm},clip,height=\imHeight]{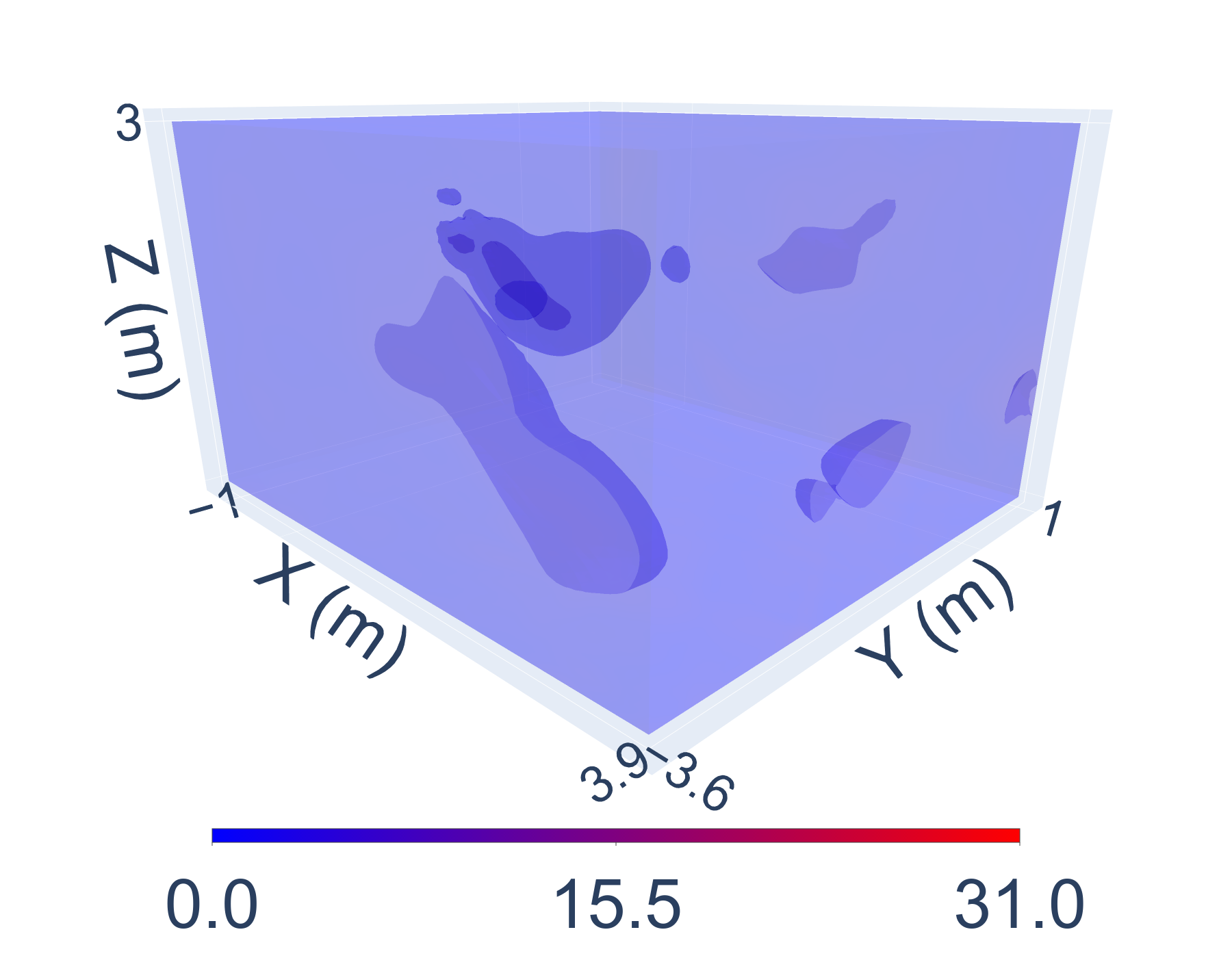}
        \label{fig:bos-err}
    \end{subfigure}
    \hfill
    \begin{subfigure}{0.2\linewidth}
        \centering
        \caption*{ ~}
        \includegraphics[trim={50mm 5mm 50mm 50mm},clip,height=\imHeight]{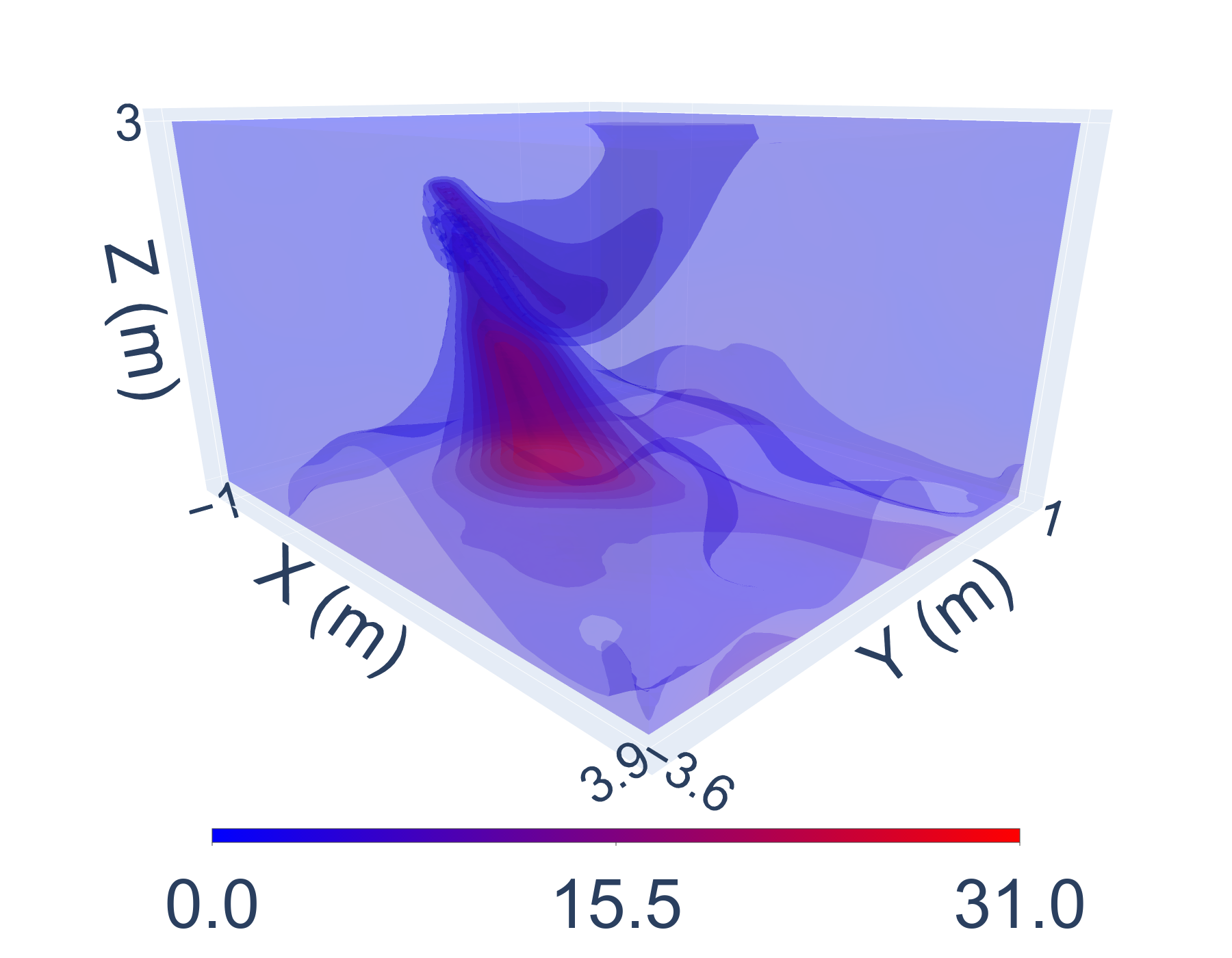}
        \label{fig:pde-bound-err}
    \end{subfigure}
    \hfill
    \begin{subfigure}{0.2\linewidth}
        \centering
        \caption*{ ~}
        \includegraphics[trim={50mm 5mm 50mm 50mm},clip,height=\imHeight]{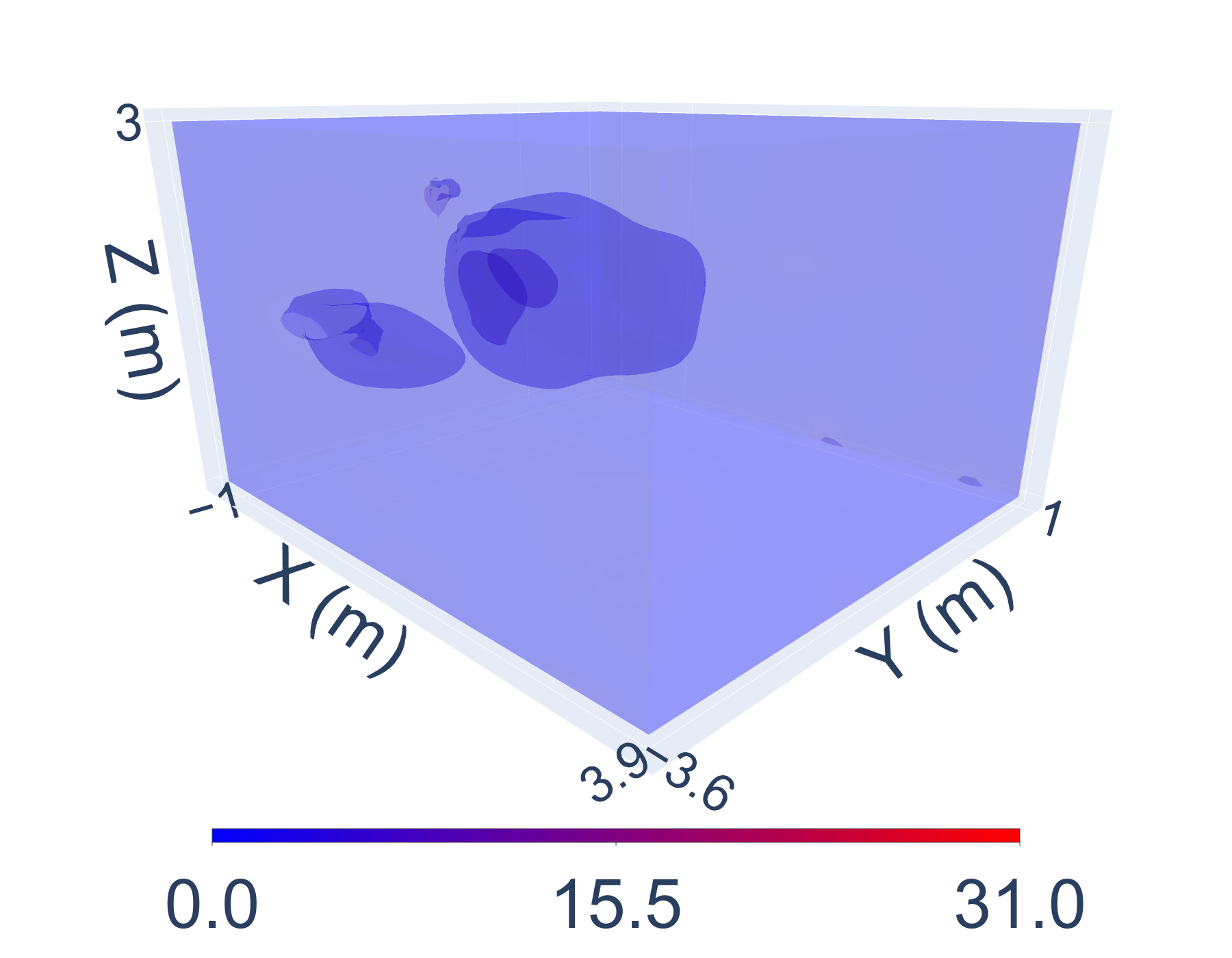}
        \label{fig:bos-pde-bound-err}
    \end{subfigure}
    \vspace{-0.2in}
    \caption{
    \textit{Airflow reconstruction results using the projector and camera BOS acquisition setup. Top row from left to right shows the ground truth and reconstructions of the temperature field (in $\unit{\kelvin}$) using different combinations of the losses. Bottom left figure illustrates the BOS image measurement. The remaining bottom row plots illustrate the absolute error of each reconstruction regime.}} \vspace{-0.2in}
    \label{fig:results}
\end{figure*}

\subsection{Physics-Informed BOS Tomography}
Reconstructing the temperature $\fluidt$, pressure $\fluidp$, and velocity $\fluidu$ fields of the airflow can be formulated as a tomographic inverse problem given BOS image measurements $I_{\rm flow}$ and boundary conditions $(\fluidt^*, \fluidp^*, \fluidu^*)\big|_{\Gamma}$ at a boundary region $\Gamma$. Single-view 3D BOS has inherent ambiguities along the view direction~\cite{zhao_single_2024}. Following the PINNs framework~\cite{raissi_physics-informed_2019, wang_experts_2023}, we propose to regularize the inversion using a physics-informed loss. This is done by solving the following optimization problem:
\begin{equation}\label{eq:inverse_problem}
\begin{array}{ll}
    \min\limits_{\fluidt, \fluidp, \fluidu} &\lambda_1\loss_{\rm BOS}(\eta) + \lambda_2\loss_{\Gamma}(\fluidt, \fluidp, \fluidu) + \lambda_3 \loss_{\rm PDE}(\fluidt, \fluidp, \fluidu)\\
    \textrm{subject to} & \eta = 1 + \rho_0 G\frac{T_0}{T},
\end{array}
\end{equation}
This objective function includes the following loss components:
\begin{itemize}[leftmargin=*, itemsep=1mm, topsep=0mm]
\item The BOS image loss $\mathcal{L}_{\rm BOS}= \sum
    _j \Bigl\|I^j_{\rm flow} - I^j(\eta)\Bigr\|_2^2$,
    where $j$ denotes the pixel index, $I_{\rm flow}$ is the BOS image measurement intensity, and $I(\eta)$ is the predicted intensity from~\eqref{eqn:imageformation}.
\item The boundary loss $\mathcal{L}_{\Gamma} = \left\|(\fluidt^*_{\rm n}, \fluidp^*_{\rm n}, \fluidu^*_{\rm n})\big|_{\Gamma} - (\fluidt_{\rm n}, \fluidp_{\rm n}, \fluidu_{\rm n})\big|_{\Gamma}\right\|_2^2$,
where subscript ${\rm n}$ denotes the field divided by its maximum value.
\item The physics-informed loss $
    \mathcal{L}_{\rm PDE} =  \sum_{i=1}^{N_{\rm c}} \gamma_1 r_{\text{mass}}^2\left(\fluidx_i\right) + \gamma_2 \| \mathbf{r}_{\text{mom}}\left(\fluidx_i\right)\|_2^2 + \gamma_3 r_{\text{heat}}^2\left(\fluidx_i\right)$,
where $\gamma_{1,2,3}$ are scalar multipliers that balance the weight of each residual, and $\fluidx_{i=1,\ldots,N_{\rm c}}$ are collocation points uniformly sampled in the computational domain. $r_{\text{mass}}$, $\mathbf{r}_{\text{mom}}$, and $r_{\text{heat}}$ are the nondimensional mass conservation, momentum conservation, and heat transfer equation residuals defined by the Boussinesq approximation for buoyancy-driven flows:
\begin{align*}
&r_{\text{mass}}\left(\fluidx\right) = \nabla \cdot  \fluidu\,, \\
&\mathbf{r}_{\text{mom}}\left(\fluidx\right) = \left(\fluidu \cdot \nabla\right) \fluidu + \nabla \fluidp - \frac{1}{Re} \nabla^2 \fluidu + Ri\, T_{\rm nd}\mathbf{e}_g\,, \\
&r_{\text{heat}}\left(\fluidx\right) = \left(\fluidu \cdot \nabla\right) \fluidt_{\rm nd} - \frac{1}{Pe} \nabla^2 \fluidt_{\rm nd}\,.
\end{align*}

\end{itemize}
Here, the nondimensional temperature fluctuation $\fluidt_{\rm nd}$ is obtained from the inlet $T_{\rm in}$ and reference $T_0$ temperatures as $\fluidt_{\rm nd} = \frac{T - T_0}{T_{\rm in} - T_0}$. Additional parameters include the acceleration due to gravity $g$ and its unit vector $\mathbf{e}_g$, kinematic viscosity $\nu$, coefficient of thermal diffusivity $\alpha$, coefficient of thermal expansion $\beta$, and characteristic length $L$ and velocity $U$ scales  leading to the nondimensional Reynolds, P\'eclet, and Richardson numbers:
$
Re = \frac{ U L}{\nu} \,, Pe = \frac{U L}{\alpha} \,, Ri = \frac{g \beta \left(T_{\rm in} - T_0\right)L}{U^2}\,.
$

We use an implicit neural representation to parameterize the $\fluidt$, $\fluidp$, and $\fluidu$ fields using a multiplayer perception (MLP), i.e. $\left( \fluidt, \fluidp, \fluidu \right) = \text{MLP}\left(\pos; \theta\right)$ where $\pos$ denotes the spatial location, and $\theta$ are the MLP parameters to be determined by solving the optimization problem~\eqref{eq:inverse_problem}. 

Our forward rendering and optimization framework is implemented in JAX and uses the Equinox, Diffrax, and Optax libraries~\cite{jax2018github, kidger2021equinox, kidger2021on, hessel2010optax}. It is end-to-end differentiable where efficient gradient computation is achieved using automatic differentiation and the adjoint state method~\cite{teh_adjoint_2022, zhao_single_2024}.
\section{Experimental Setup and Results}

\subsection{Airflow simulation}
We obtain the ground truth airflow by performing a Reynolds-Averaged Navier-Stokes (RANS) simulation using OpenFOAM. The room setup and ground truth temperature field are shown in Fig.\ref{fig:gt} with an inlet pointing in the positive $x$ direction. In order to image the flow in the entire room, we simulate a hypothetical camera and projector pointing in the positive $y$ direction with a large focal length that are located at $(0.85, -29, 1.5)$m and $(1.85, -29, 1.5)$m, respectively. The projector illuminates the back-wall with a wavelet noise pattern~\cite{cook_wavelet_noise_2005}. The BOS image measurement shown in Fig.~\ref{fig:bos_meas} is obtained using our physically-based renderer for a $100\times100$ sensor resolution with 2 samples per pixel.

\subsection{Airflow reconstruction} 
 We evaluate the reconstruction performance on a $64\times64\times64$ voxelized grid given the BOS image measurement and the boundary $\fluidt, \fluidp$ and $\fluidu$ fields on the boundary $y-z$ plane at  $x = -1.2$\,m containing the inlet/outlet. Note that there is a model and resolution mismatch between the incompressible flow equations imposed by the physics-informed loss and those of the OpenFOAM RANS simulation. We run $80,000$ iterations of the Optax adabelief optimizer with mini-batch updates using $8192$ spatial points for the PDE loss, $5000$ pixels for the BOS loss, and $4096$ points for the boundary loss.

We compare the performance of reconstructing the airflow volume using our proposed BOS+PDE+boundary losses with two alternative reconstruction regimes: BOS+boundary and PDE+boundary. The reconstruction results in Fig.~\ref{fig:results} show that combining all three loss terms significantly reduces artifacts and is essential for accurately reconstructing the temperature field. We also show in~\cite{teh2025IndoorPub} that the root mean squared error is 2 order of magnitudes lower for the $\fluidp$ field and 1 order of magnitude lower for the $\fluidu$ field in the BOS+PDE+boundary reconstruction when compared with the BOS+boundary results. These findings highlight the advantages of our physics-informed and differentiable rendering framework in achieving high-accuracy BOS reconstructions of turbulent airflow.

\bibliographystyle{IEEEtran}
\bibliography{references}
\end{document}